\documentclass[10pt,a4paper,twoside]{article}

\input{rec.dat}

\usepackage{graphicx}
\usepackage{fancyhdr}
\usepackage{amsmath}
\usepackage{amssymb}
\usepackage{listings}
\usepackage{xcolor}
\usepackage{blindtext}
\usepackage{tikz} 

\definecolor{codegreen}{rgb}{0,0.6,0}
\definecolor{commentgreen}{rgb}{0,0.6,0}
\definecolor{codegreen}{rgb}{0,0.6,0}
\definecolor{codegray}{rgb}{0.5,0.5,0.5}
\definecolor{commentgray}{rgb}{0.5,0.5,0.5}
\definecolor{codepurple}{rgb}{0.58,0,0.82}
\definecolor{backcolour}{rgb}{0.95,0.95,0.92}
\definecolor{backcolour2}{rgb}{0.98,0.98,0.97}

\lstdefinestyle{mystyle}{
    backgroundcolor=\color{backcolour2},
    commentstyle=\color{commentgray},
    keywordstyle=\color{magenta},
    numberstyle=\tiny\color{codegray},
    stringstyle=\color{codepurple},
    basicstyle=\ttfamily\footnotesize,
    breakatwhitespace=false,
    breaklines=true,
    captionpos=b,
    keepspaces=true,
    numbers=left,
    numbersep=5pt,
    showspaces=false,
    showstringspaces=false,
    showtabs=false,
    tabsize=2
}

\newcommand{\inter}[1]{\overline{\underline{#1}}}
\newcommand{\interx}{\overline{\underline{x}}}
\newcommand{\interz}{\overline{\underline{z}}}

\newcommand{\lo}[1]{\underline{#1}}
\newcommand{\hi}[1]{\overline{#1}}

\begin{document}

\title{\TitleFont Interval propagation through the discrete Fourier transform}




 \author[1*]{Marco De Angelis\authorsep}
 \author[1,2]{Marco Behrendt\authorsep}
 \author[1]{Liam Comerford\authorsep}
 \author[3]{Yuanjin Zhang\authorsep}
 \author[1,2]{Michael Beer \lastauthorsep}
 \affil[1]{Institute for Risk and Uncertainty, University of Liverpool, United Kingdom}
 \affil[2]{Institute for Risk and Reliability, Leibniz University Hannover, Germany}
 \affil[3]{School of Safety Science and Emergency Management, Wuhan University of Technology, China}
 \affil[*]{\corremail{marco.de-angelis[at]liverpool.ac.uk} }

\begin{abstract}
\noindent
We present an algorithm for the forward propagation of intervals through the discrete Fourier transform. The algorithm yields best-possible bounds when computing the amplitude of the Fourier transform for real and complex valued sequences.
We show that computing the exact bounds of the amplitude can be achieved with an exhaustive examination of all possible corners of the interval domain. However, because the number of corners increases exponentially with the number of intervals, such method is infeasible for large interval signals. We provide an algorithm that does not need such an exhaustive search, and show that the best possible bounds  can be obtained propagating complex pairs only from the convex hull of endpoints at each term of the Fourier series. Because the convex hull is always tightly inscribed in the respective rigorous bounding box resulting from interval arithmetic, we conclude that the obtained bounds are guaranteed to enclose the true values.


\end{abstract}

\maketitle
\thispagestyle{titlestyle}

\section{Introduction}\label{sec:intro}

The analysis of engineering structures subject to harmonic stochastic excitation is often conveniently done in the Fourier domain \cite{newland1993introduction}. The response of the structure is obtained for each harmonic without the need for step-wise numerical integration. This kind of analysis however, seems to be incompatible with the fact that real data time-series records are often affected by uncertainty. Some authors have successfully propagated interval uncertainty in dynamic structural analysis directly in the time domain \cite{muscolino2012frequency}. In other instances the literature on the topic seem to contemplate only purely probabilistic methods to address the problem of characterization of the uncertainty of the power spectrum from real data, as shown in  \cite{comerford2015quantifying} and literature therein.
It is currently very difficult to convert an imprecise time signal to its Fourier domain. An algorithm for the interval Fourier transform is much needed to bridge the divide between imprecise signals and the well established harmonic structural analysis, and to enlarge the spectrum of engineering applications that explicitly account for uncertainty.

The discrete Fourier transform (DFT) is used in a variety of different applications in science and engineering, such as in spectral analysis, random vibration, differential equations, data compression, signal processing, image processing, probabilistic programming and many more \cite{sneddon1995fourier,james2011studentsguide,newland1993introduction}. There are various algorithms available for transforming a signal with the DFT, of which the best known is probably the fast Fourier transfom (FFT), presented by Cooley \& Tukey \cite{cooley1965algorithm}. Due to increasing computational power, simulations and equivalent calculations can be carried out ever faster, which is particularly important for the numerical analysis. An overview of a variety of algorithms used can be found in abundance in \cite{cormen2009introduction,stoer2013introduction}.

Real data records are required for a large number of problems in engineering. A major challenge with data acquisition that should not be neglected is that sensors only work accurately within certain tolerances and are therefore subject to uncertainties. These uncertainties arise, for instance, due to damaged sensors, device failures and measurement errors, or if the sensors are not precisely calibrated. In addition, the data could be captured incorrectly due to threshold limitations. For more accurate simulation results when utilising real data records, uncertainties must be taken into account \cite{comerford2015quantifying}.

In this work, three different methods are investigated to provide bounds on the amplitude of the DFT: (1) a brute-force method, (2) a method based on the convex hull, (3) a method based on complex interval arithmetic and rigorous bounding \cite{alefeld2012introduction,moore1966interval,moore1979methods,moore2009introduction}. These methods provide a direct relationship between interval signals and the corresponding amplitudes of the DFT. Thus, intervals can replace the exact values when investigating structures while the signal's uncertainty can be captured by these intervals. 

\section{Problem statement}
In this paper the focus is on interval-valued  real signals denoted by $\inter{x}_n$ \cite{alefeld2012introduction}. An interval-valued real number, or real interval, or simply an \emph{interval} is a non-empty, closed, and bounded subset of the real line $\mathbb{R}$,

$$\inter{x} := [\lo{x}, \hi{x}] := \{ x \in \mathbb{R} | \lo{x}  \leq x \leq \hi{x} \}.$$

Similarly a \emph{complex interval} denoted by $\interz$, can be defined with a pair of intervals, one for the real component $\inter{z}_{\text{re}}$ and the other for the imaginary component $\inter{z}_{\text{im}}$,

$$\inter{z} := \inter{z}_{\text{re}}  + i ~ \inter{z}_{\text{im}} := \{ z_{\text{re}} + i~ z_{\text{im}} ~|~ z_{\text{re}} \in \inter{z}_{\text{re}} \subseteq \mathbb{R} ~  \& ~  z_{\text{im}} \in \inter{z}_{\text{im}} \subseteq \mathbb{R}  \},$$

for more about the space of complex intervals see \cite{alefeld2012introduction}. The interval extension of the DFT converts an interval signal (or ordered sequence) $ \interx_0, \interx_1, ..., \interx_{n}$ with $n<N$ , into the respective signal of interval complex numbers $\interz_0, \interz_1, ..., \interz_{k}$, with $k<N$, also known as Fourier sequence

\begin{equation}\label{eq:intervalDFT}
\inter{z}_k := \sum_{n=0}^{N-1} \interx_n \ e^{-\frac {i 2\pi}{N}kn}  =   \sum_{n=0}^{N-1} \interx_n \ \left( \cos \frac{2 \pi}{N}kn - i \ \sin \frac{2 \pi}{N}kn \right).
\end{equation}

When $x$ is a time signal, each complex number $z_k \in \mathbb{C}$ represents an harmonic of the signal with angular frequency $2\pi k n /N$. The DFT function is sometimes denoted by $\mathcal{F}$, while its interval extension by $ [\mathcal {F}]$, where $\interz_k =  [\mathcal {F}]\left(\interx_k \right)$. 

\begin{figure}[htb] 
\centering
\includegraphics[width=1 \textwidth]{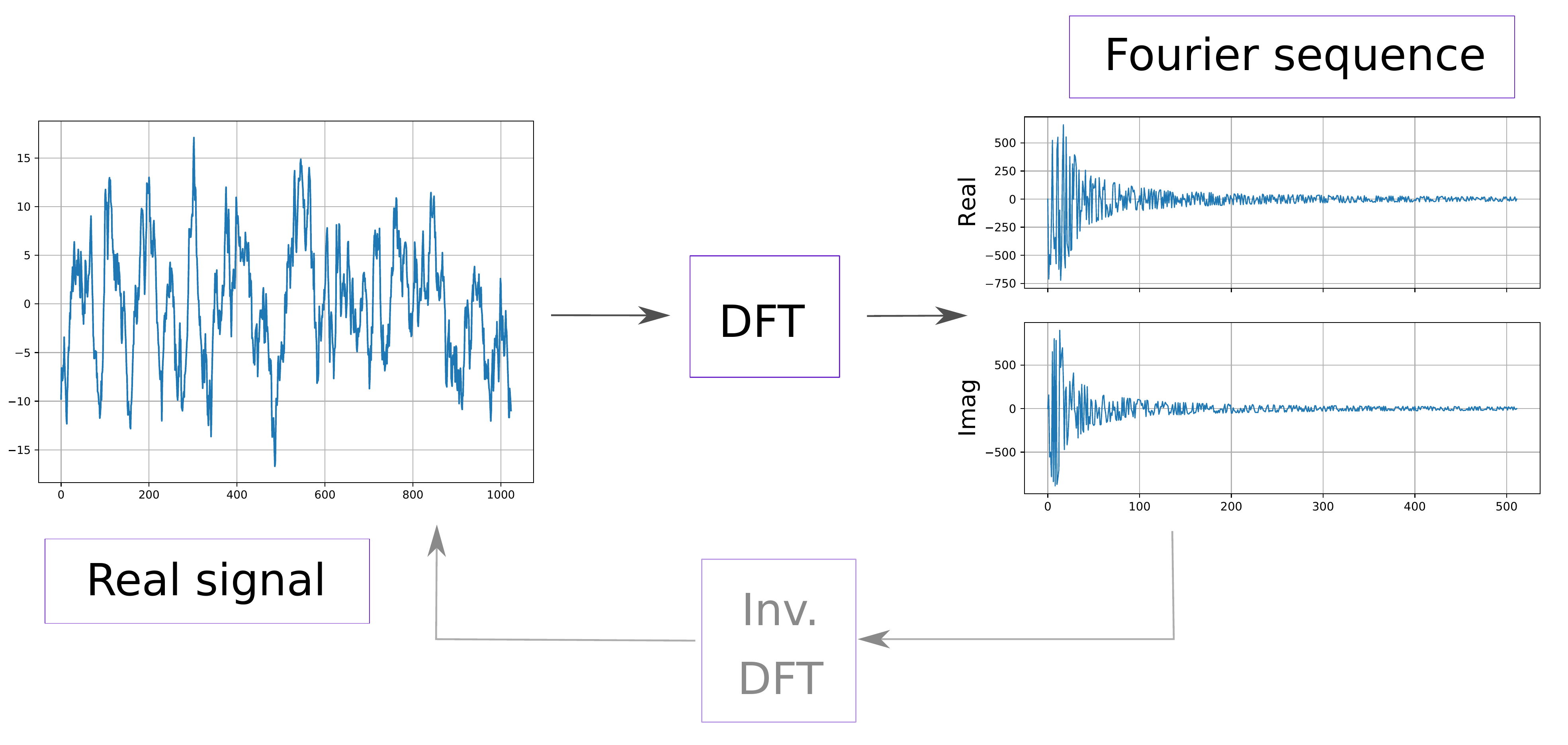}\\
\caption[]{Discrete Fourier transform of a real signal to its Fourier sequence.} \label{fig:DFT_diagram}
\end{figure}

\begin{figure}[h!] 
\centering
\includegraphics[width=1 \textwidth]{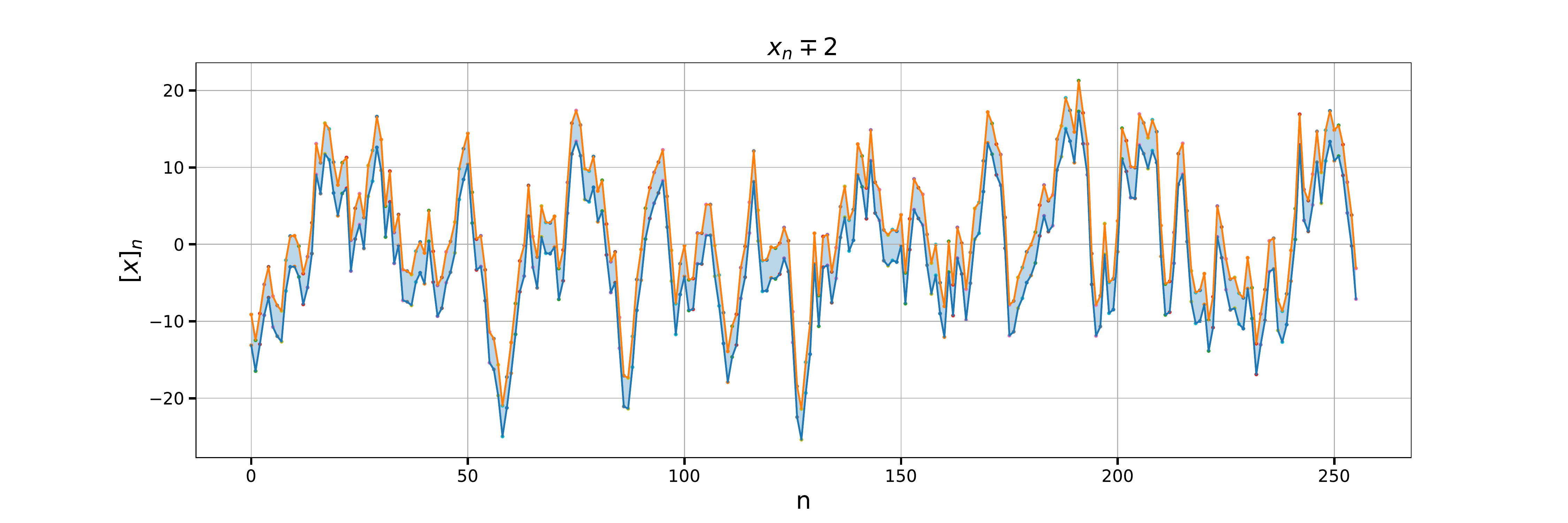}\\
\caption[]{An interval signal $ \interx_0, \interx_1, ..., \interx_{n}$, with N=128 points, and interval uncertainty $\mp 2$.} \label{fig:intervalsignal}
\end{figure}

The DFT in practice is used to convert a time signal into its Fourier sequence (Figure \ref{fig:DFT_diagram}) to study the harmonic properties of the signal where each element corresponds to a frequency. The DFT is often used to reduce the dimensionality of the signal; the inverse Fourier transform is used to reconstruct the signal after compression or simply to convert the signal into the time domain.

The objective of this work is to provide best-possible bounds on the \emph{amplitude} of $\mathcal {F}$, a.k.a. amplitude spectrum, 

\begin{equation}\label{eq:amplitude_IDFT}
\text{abs} \left( \interz_k \right) := \left| \interz_k \right| := \sqrt{  \left( \sum_{n=0}^{N-1}   \interx_n  \  \cos \frac{2 \pi}{N}kn \right)^2 + \left( \sum_{n=0}^{N-1} \interx_n  \   \sin \frac{2 \pi}{N}kn \right)^2  }.
\end{equation}

The interval signal of Figure \ref{fig:intervalsignal} will be considered throughout the manuscript. 
A convenient way to evaluate the interval extension $[\mathcal{F}]$ is to split non-interval and interval components, as shown in  (\ref{eq:intervalDFT_sameprecision}).
For example, when the signal is acquired using the same sensor (same precision) the interval uncertainty has the same value throughout. In such case the $[\mathcal {F}]$ becomes

\begin{equation}\label{eq:intervalDFT_sameprecision}
\inter{z}_k := \sum_{n=0}^{N-1} x_n \ e^{-\frac {i 2\pi}{N}kn}  +  \sum_{n=0}^{N-1} \xi\ \Delta \ e^{-\frac {i 2\pi}{N}kn},
\end{equation}

where $\Delta = [-1,\ 1]$ is a unitary interval, and $\xi \in \mathbb{R}$ is the precision of the interval signal, expressed in the same units as the signal. 
This form can be used to assess the effect of the cumulation of uncertainty on the amplitude as more and more uncertainty components get added. Also the non-interval term can be moved to the left-hand side of the equal sign of (\ref{eq:intervalDFT_sameprecision}) to yield a centred form of the interval Fourier transform.

\section{Interval arithmetic}

The interval extensions defined in  (\ref{eq:intervalDFT}), (\ref{eq:amplitude_IDFT}), and (\ref{eq:intervalDFT_sameprecision}) can be evaluated with interval arithmetic. The generalized rules of arithmetic for \emph{addition}, \emph{multiplication} and \emph{power elevation} are needed for the purpose. 
The addition $+$ between any two intervals $\inter{x},\  \inter{y}$ with  $ x,\ y  \in \mathbb{R}$ is 
    \begin{equation}
       \inter{x} + \inter{y} := [\lo{x}+\lo{y},\ \hi{x}+\hi{y}]. 
    \end{equation}\label{eq:sum_interval}
The multiplication between an interval and a number $a \in \mathbb{R}$, which symbol $\times$ is often omitted is
    
    \begin{equation}
        a\ \interx := 
        \left\{\begin{matrix}
          [a  \lo{x},\ a \hi{x}], \  a > 0; \\
          [a \hi{x},\ a \lo{x}], \  a < 0.
        \end{matrix}\right.
    \end{equation}\label{eq:atom_product}
The power elevation with exponent two is

 \begin{equation}
        \interx^2 := \left\{ \begin{matrix}
          [\lo{x}^2,\ \hi{x}^2], \ \interx \geq 0\\
          [\hi{x}^2,\ \lo{x}^2], \  \interx < 0\\
          [0, \max{ \left( \lo{x}^2,\  \hi{x}^2 \right) }, \  \interx \ni 0 .
        \end{matrix}\right.
    \end{equation}\label{eq:power2_interval}
The square root of a positive real interval is 

  \begin{equation}
        \sqrt{\interx} :=  [\sqrt{\lo{x}},\  \sqrt{\hi{x}}], \ \interx > 0.
    \end{equation}\label{eq:sqrt_interval}

With these rules the amplitude of the DFT interval extension can be evaluated using (\ref{eq:amplitude_IDFT}) without the need of a particular algorithm. The results provided by the interval arithmetic are guaranteed to enclose the exact bounds. However artefactual uncertainty due to the dependence problem makes the bounds obtained with interval arithmetic wider than they ought to be.

The interval extension of the DFT (\ref{eq:intervalDFT}) can be computed using the complex addition operator. Consider two complex intervals $\inter{z}_k  = [\lo{z}_k, \hi{z}_k], \ k=1,2$, their addition is given by 

$$\inter{z}_1+\inter{z}_2 := [\lo{z}_1+\lo{z}_2, \hi{z}_1+\hi{z}_2].$$

With the complex addition operator the program used to compute $\mathcal {F}$ can be recycled to obtain its interval extension. The input \emph{signal} to the program will be an ordered sequence of intervals.

\begin{lstlisting}[language=Python, caption={Code sample for a basic Python implementation of the Fourier transform.}, captionpos=b, label={lst:dft}]
def DFT(signal):  # Inputs an interval signal
    F=[] 
    N = len(signal)  # length of the real signal 
    for k in range(N//2):  # for each frequency, with k=0,1,2,...
        f = 0
        for n in range(N):  
            f += signal[n]*exp(-1j*2*pi*k*n/N)  
        F.append(f)
    return F
\end{lstlisting}

Code sample Listing \ref{lst:dft} can be used independently from external interval libraries, replacing an interval with an ordered list of a complex pair. The function in Listing \ref{lst:dft} do need $numpy$ for $\pi$ and $\exp$.
The amplitude of $\mathcal {F}\left(\mathbf {\interx} \right)$ for each frequency  $\omega_k,\ k=0,1,2,...$ can be obtained replacing $\text{f}$ at line 8 with $\text{abs}(\text{f})$, or with the following function.

\begin{lstlisting}[language=Python, caption={Code sample for the DFT amplitude.},captionpos=b, label={lst:amplitude}]
def DFT_amplitude(signal): 
    F = DFT(signal)
    return [abs(f) for f in F]
\end{lstlisting}

The only caveat in Listing \ref{lst:dft}, is that the length of the signal must be a power of two in order for the signal to be fully transformed to the Fourier domain. Slightly modified programs can be obtained with little extra effort to map signals that are not a power of two, see e.g. \cite{newland1993introduction}.

\subsection{Notes on the dependence problem using interval arithmetic}
The bounds on the amplitude of the interval extension $\mathcal {F}$  evaluated with interval arithmetic are \emph{rigorous} in the sense that all the possible signals $\mathbf{x} \in \mathbf{\interx}$ are mapped inside these bounds. However when the amplitude is computed, the bounds obtained with (\ref{eq:amplitude_IDFT}) will be suboptimal because of the dependence problem. The computation of  (\ref{eq:amplitude_IDFT}) with interval arithmetic does not track the dependence between the real and the imaginary component, thus the bounds are outer-estimated. The bounds are rigorous because all of the possible dependencies between real and imaginary components are considered by the interval arithmetic bounds.

\section{The brute-force method}

The \emph{brute-force} method is too inefficient to be used in practical applications, and it is herein used mainly for illustration purposes. The method tracks the dependency between real and imaginary components by constructing a binary tree of all the propagated endpoints at each iteration step. Clearly the number of endpoints increases exponentially in base two and with exponent given by the iteration number. The total number of iterations is the length of the signal, so the longer the signal the more inefficient is the \emph{brute-force} method.  Thus this method can only be used for research purposes either on a very short signal or on a signal with very few intervals. On a personal computer the recommended size of the problem is eight, more iterations may slow down the process significantly.
 
\begin{lstlisting}[language=Python, caption={Code sample for Python implementation of the \emph{brute-force} method},captionpos=b, label={lst:bruteforce}]
def BruteForce(intervalsignal, k, Limit): 
    N = len(intervalsignal)  # <- length of the signal
    COMB=[]
    pair = [complex(exp(-2*pi*1j*k*0/N)) * iend for iend in intervalsignal[0]] 
    pairs = pair  # initialise set of endpoints
    COMB.append(pairs)
    for n in range(1,Limit): # Limit < N when N is large
        pair  =  [complex(exp(-2*pi*1j*k*n/N)) * iend for iend in intervalsignal[n]] 
        pairs = [[pair[0] + ps for ps in pairs],[pair[1] + ps for ps in pairs]] 
        COMB.append(pairs) # add to list of endpoints 
        pairs = pairs[0] + pairs[1] # update for next iteration
    return COMB
\end{lstlisting}

\begin{figure}[ht!] 
\centering
\includegraphics[width= \textwidth]{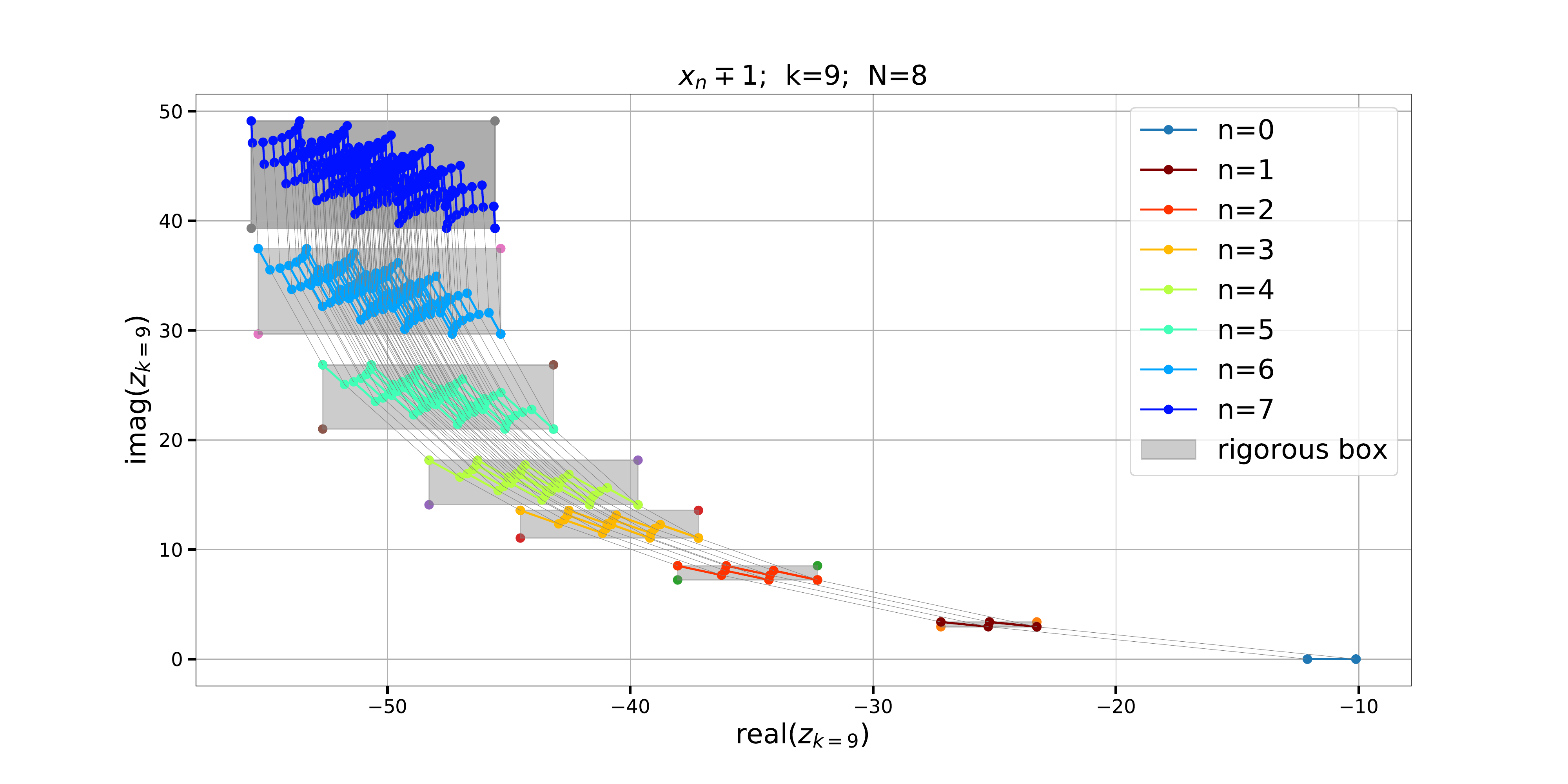}\\
\caption[]{Set of endpoints produced by the \emph{brute-force} method v. rigorous box.} \label{fig:interval_box_k9}
\end{figure}

Because the \emph{brute-force} method is so inefficient, we do not provide the code for computing the bounds out of the set of endpoints outputted by the program. The code in Listing \ref{lst:bruteforce} can be used without external interval libraries; however it does require \emph{numpy} for $\pi$ and $\exp$. The program \ref{lst:bruteforce} outputs the sets of points depicted in Figure \ref{fig:interval_box_k9}, which shows that the bounds provided by interval arithmetic in the complex plane are rigorous and best possible. The Figure superimposes the propagated endpoints through the Fourier transform on the interval arithmetic box, for a given \emph{frequency} ($k=9$) of the Fourier domain, and up to eight iterations. This is because there are no repeated variables in the complex expression of (\ref{eq:intervalDFT}). 
Because the DFT is a linear function, mapping an \emph{interval} signal to its Fourier domain requires projecting the endpoints only. 
Because there are no repeated variables, the number of interval components present in the signal does not affect the inflation of the bounds that keep enclosing the endpoints tightly. The final interval box, for a given frequency always tightly enclose the set of propagated endpoints. For example if the signal had eight components, given $k=9$, we would be looking at the last gray box ($n=7$).

\section{The selective method}

We observe that only the endpoints on the convex hull on the current set of endpoints carry the information about the bounds. Numerical experiments performed using the \emph{brute-force} method have confirmed this intuition, however the investigation is limited by the number of iterations allowed by the brute-force method. With interval arithmetic we can keep propagating the endpoints on the convex hull as well as the enclosing box for as many iterations as needed and for however large signals. Moreover, because interval arithmetic is the most efficient method its box can always be computed alongside the \emph{selective}  method. 

\begin{figure}[ht!] 
\centering
\includegraphics[width= \textwidth]{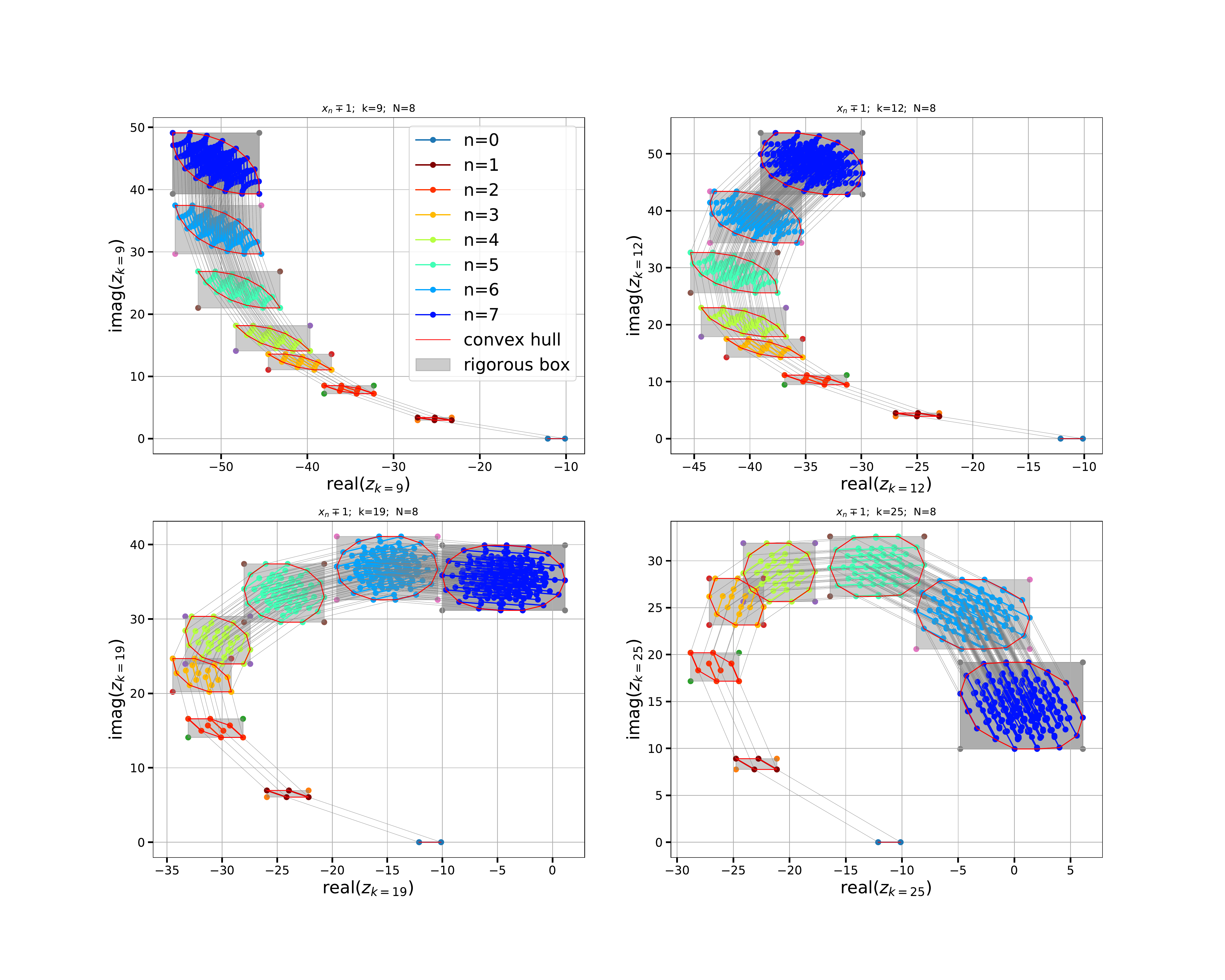}\\
\caption[]{Set of points on the convex hull and interval box for four different frequencies.} \label{fig:bf_4freq}
\end{figure}

The numerical experiments done running the \emph{selective}  method alongside interval arithmetic have confirmed and \emph{verified} the intuition that the selective method yields rigorous best possible bounds in each tested case. Figure \ref{fig:bf_4freq}, shows on four different frequencies, the three methods altogether. The convex hull is the set of points resting on the set boundary, which in Figure  \ref{fig:bf_4freq} is depicted by a red closed line. The convex hull is obtained at each iteration by propagating only the endpoints on the boundary of the previous iteration step. 

Interval arithmetic gives us the verified guarantee that the \emph{selective} method is rigorous, i.e. does not underestimate the uncertainty. The fact that the \emph{selective} method makes use of a subset of the endpoints ensures that its bounds are also best possible. In fact, it is always possible to index the convex-hull endpoints resting on the edge of the rigorous box and propagate these points back to identify the combination of endpoints yielding the bounds. However because there is no guarantee that theses endpoints will be unique, the algorithm cannot be used directly to perform the inverse transform.

The code produces the set of points in the convex hull as shown in Listing \ref{lst:selective} and can be obtained as a derivation from the \emph{brute-force} method. The function \emph{Selective} in the code snippet Listing \ref{lst:selective} will be invoked for each frequency $k$ of the Fourier sequence. 

\begin{lstlisting}[language=Python, caption={Code snippet for the Python implementation of the \emph{selective} method.},captionpos=b, label={lst:selective}]
from scipy.spatial import ConvexHull
def Selective(intervalsignal, k): 
    N = len(intervalsignal)
    CHULL = []
    pair = [complex(exp(-2*pi*1j*k*0/N)) * iend for iend in intervalsignal[0]] 
    chullpair = pair
    CHULL.append(chullpair)
    for n in range(1,N):
        pair =  [complex(exp(-2*pi*1j*k*0/N)) * iend for iend in intervalsignal[n]] 
        chullpair=[[pair[0]+ch for ch in chullpair],[pair[1]+ch for ch in chullpair]]
        chullpair = chullpair[0] + chullpair[1]
        chullpairs_RI = [[p.real, p.imag] for p in chullpair]
        hull = ConvexHull(chullpairs_RI)
        chullpair = [chullpairs_RI[h][0] + 1j*chullpairs_RI[h][1] for h in hull.vertices]
        CHULL.append(chullpair)
    return CHULL
\end{lstlisting}

\section{The bounding-box method}

There are no repeated variables in the interval extension of (\ref{eq:intervalDFT}), thus the final enclosing box is the smallest box possible containing the set of points generated by the \emph{brute force} method. In Listing \ref{lst:interval} we show the procedure to obtain the enclosing interval box at each iteration of the Fourier transform. Note that if the enclosing box was not the smallest or \emph{best possible}, there would have been a gap between the box and the convex hull of points, effectively making it impossible for us to verify the results produced by the selective method. Luckily in our numerical experiments there has been no instance of such a case. The rigorous bounding method can be performed alongside the selective method with very little additional computations, to provide a prompt verification certificate of the bounds outputted by the selective method.

\begin{lstlisting}[language=Python, caption={Code snippet for the Python implementation of the \emph{interval} method.},captionpos=b, label={lst:interval}]
def RigorousBox(intervalsignal, k): 
    N = len(intervalsignal)
    RBOX = []
    firstinterval = [complex(exp(-2*pi*1j*k*0/N)) * iend for iend in intervalsignal[0]] 
    RBOX.append(firstinterval)
    ci =firstinterval
    for n in range(1,N):
        ci += complex(exp(-2*pi*1j*k*n/N)) * intervalsignal[k]
        RBOX.append(ci)
    return RBOX
\end{lstlisting}

The Fourier amplitude spectrum can be obtained just using the information carried by the corners of the enclosing box. Because the set of extreme endpoints rest on the convex hull, the amplitude obtained with the enclosing box will carry inflated uncertainty. The bounds on the amplitude will be reflective of the fact that the zone outside the convex hull between  the box will be propagated too, effectively making the bounds on the amplitude much puffier. The extra puffiness produced by this method however is compensated by the efficiency of this method, which practically carries no additional computational cost when compared to the \emph{precise} DFT.

\section{The final algorithm for the bounds on the amplitude}

\begin{figure}[h!]
\centering
\begin{tikzpicture}
\node (rect) [draw, align = center] (BF) at (0,1.8) {Brute Force $O(2^N)$};

\draw[gray, very thin, fill=gray!20!, fill opacity=0.4] (-2.5,-2.5) rectangle (2.5,.5);
\node (rect) [draw, align = center] (BS) at (0,-.25) {Selective method $O(N^2 \log N)$};
\node (rect) [draw, align = center] (BI) at (0,-1.75) {Bounding-box method $O(N)$};

\node (rect) [draw, align = center] (IF) at (5,-1) {Interval \\ Fourier transform};

\draw[gray, ->, line width=0.6pt] (BF) -- (BS);
\draw[gray, ->, line width=0.6pt] (2.5,-1) -- (IF);

\node[] (f1) at (5,1.5) {\pgftext{\includegraphics[width=.3\textwidth]{./figs/intervalsignal_2.pdf}}};
\node[] (f2) at (10,-1.) {\pgftext{\includegraphics[width=.3\textwidth]{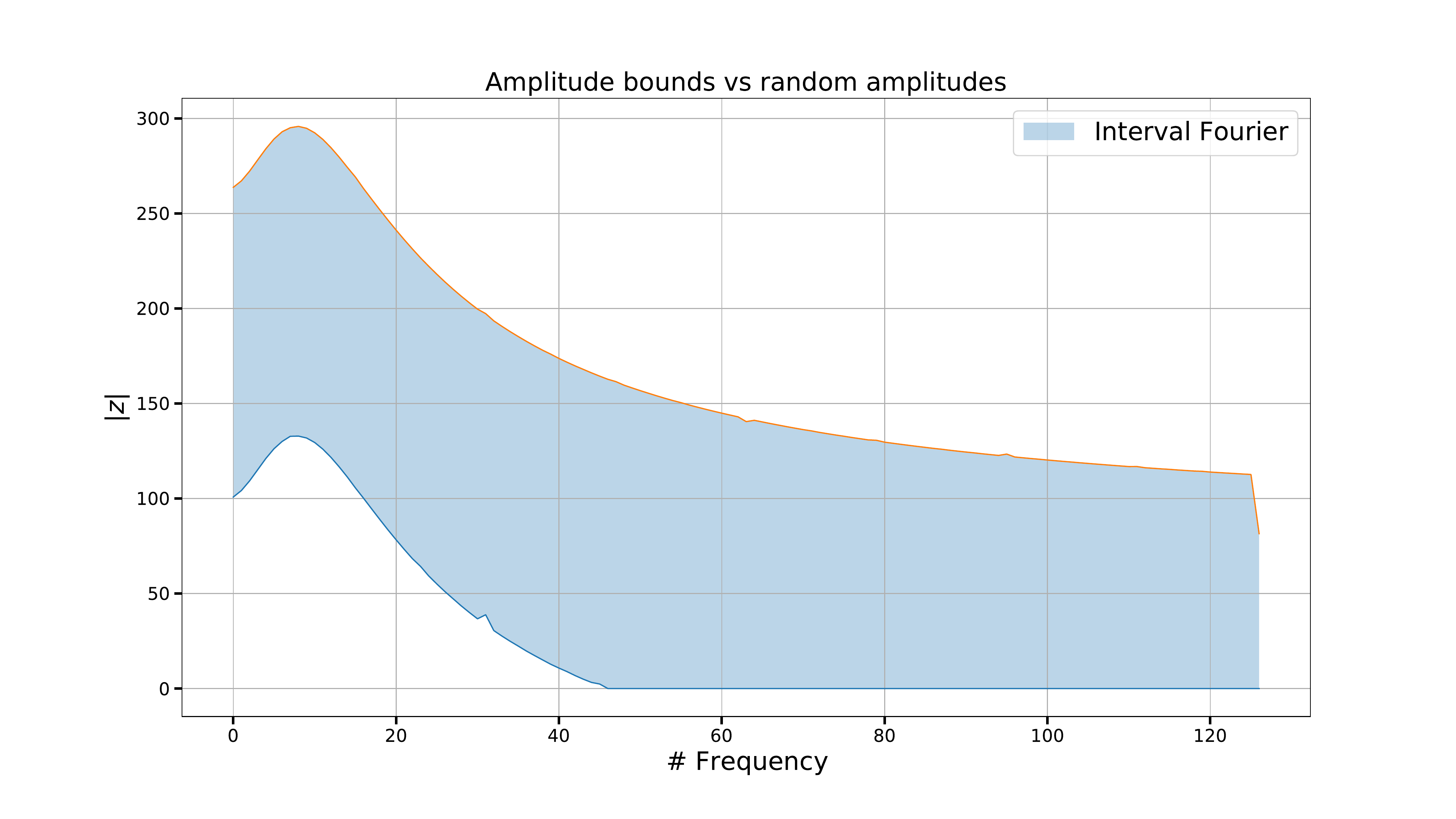}}};

\draw[gray, ->, line width=0.6pt] (f1) -- (IF);
\draw[gray, ->, line width=0.6pt] (IF) -- (8,-1);
\end{tikzpicture}\caption{Interval Fourier transform algorithm in summary.}\label{fig:chartmethods}
\end{figure}

In this section we show how to obtain the bounds on the amplitude for each frequency. A schematic representation of how the three methods interact with each other is provided in Figure \ref{fig:chartmethods}. The cost of the selective method is $O(N^2 \log N)$ because a convex hull is computed at each iteration. The divide and conquer algorithm for computing the convex hull has $O(N \log N)$ cost, hence the total cost of the selective method.

\begin{figure}[ht!] 
\centering
\includegraphics[width= \textwidth]{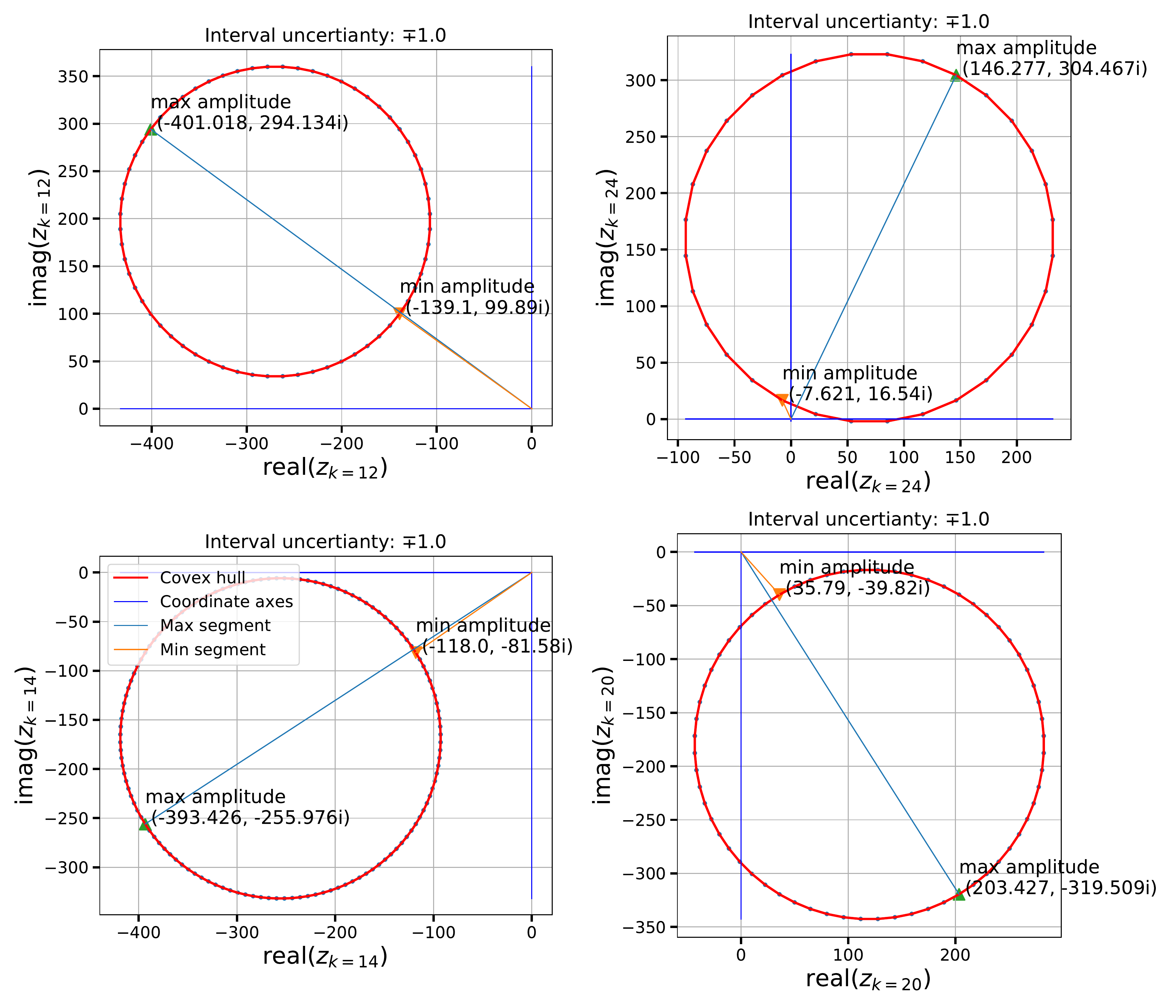}\\
\caption[]{Convex hull of endpoints for an interval signal N=128 long, and four different frequencies.} \label{fig:minmaxampl}
\end{figure}

\begin{figure}[ht!] 
\centering
\includegraphics[width= 0.5\textwidth]{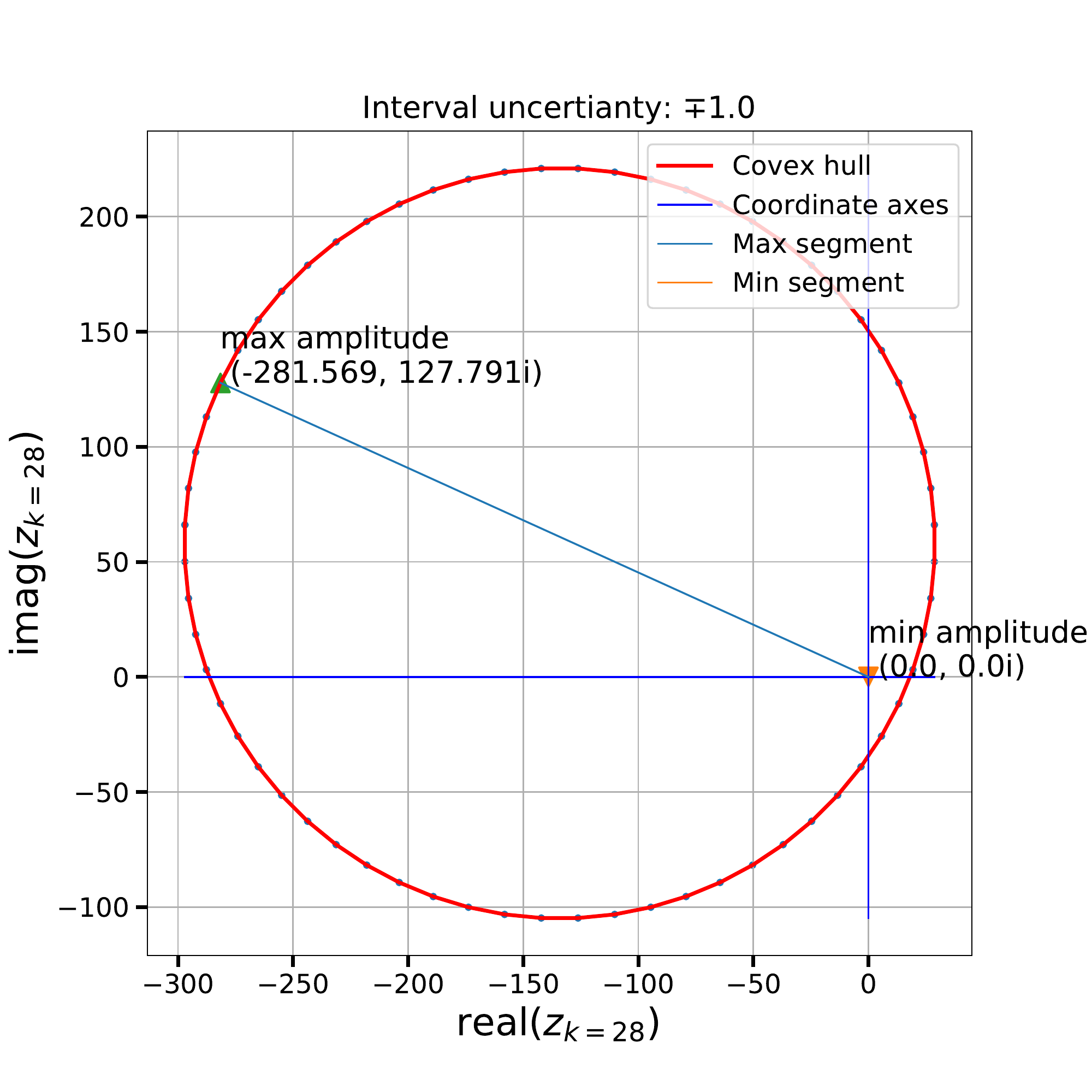}\\
\caption[]{Convex hull of endpoints enclosing the origin of the complex plain.} \label{fig:minmaxampl0}
\end{figure}

The convex hull and the interval box at the last iteration for $n=N-1$, can both be used to compute the amplitude at each frequency. We have shown that the endpoints on the convex hull yield the best possible bounds, thus the bounds on the amplitude are obtained by taking the \emph{minimum} and \emph{maximum} amplitude corresponding to this set of endpoints. The points corresponding to the \emph{extrema} are shown in Figure \ref{fig:minmaxampl}. Because the amplitude of the Fourier transform is effectively the Euclidean norm on the $\mathbb{R}^2$ isomorphism of the complex plane, minimum and maximum correspond to the points that are nearest and furthest from the origin of the complex plane. However, this is true as long as the convex hull does not enclose the origin. When the convex hull encloses the origin the \emph{minimum} will be attained at the origin itself. 
The algorithm for the amplitude bounds will therefore assume that the  \emph{minimum} is attained at the origin of the complex plane in such situation, as shown in Figure \ref{fig:minmaxampl0}.

A simple implementation of the algorithm for the amplitude bounds with the selective method is shown in Listing \ref{lst:amplitsel}. The function \emph{AmplitudeBounds} needs an external function called \emph{origin\_in\_complex\_hull} that returns \emph{True} when the convex hull encloses the origin and a \emph{False} otherwise.

\begin{lstlisting}[language=Python, caption={Python function for the amplitude bounds with the \emph{selective} method.},captionpos=b, label={lst:amplitsel}]
def AmplitudeBounds(intervalsignal, k): 
    CHULL = Selective(intervalsignal, k)
    convhull = CHULL[-1]
    chull_max = numpy.argmax([abs(h) for h in convhull])
    chull_min = numpy.argmin([abs(h) for h in convhull])
    if origin_in_complex_hull(convhull):
        return 0, abs(convhull[chull_max])
    else:
        return abs(convhull[chull_min]), abs(convhull[chull_max])
\end{lstlisting}

\begin{figure}[h!] 
\centering
\includegraphics[width= \textwidth]{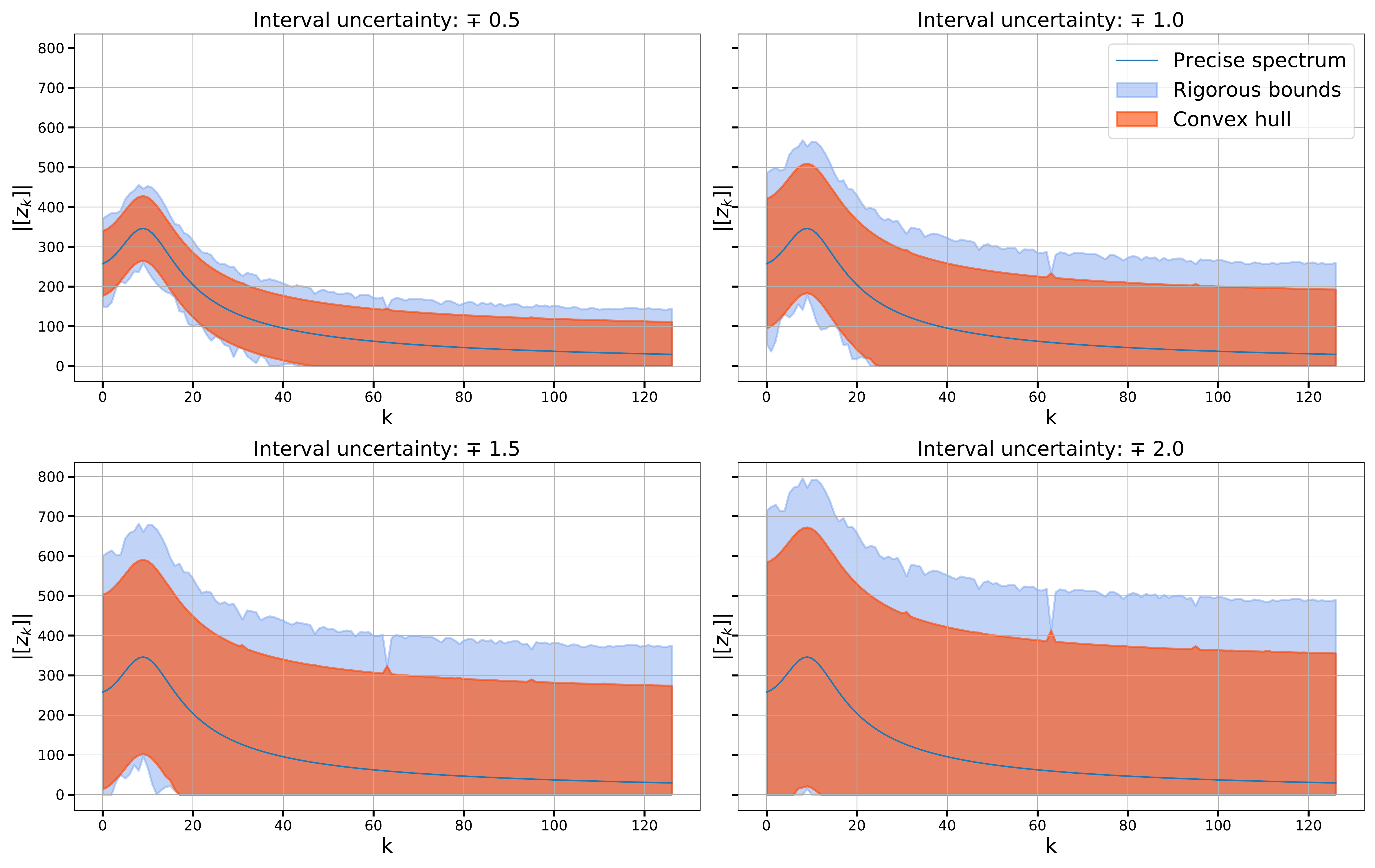}\\
\caption[]{Amplitude bounds for the Fourier transform for different uncertainty levels.} \label{fig:allfreq}
\end{figure}

The procedure for the amplitude spectrum bounds with the interval box is equivalent except that is much easier to check if the origin is contained in the box. Iterating over all the frequencies of the Fourier sequence as shown in Listing \ref{lst:allfreq} leads to the amplitude bounds shown in Figure \ref{fig:allfreq}.

\begin{lstlisting}[language=Python, caption={Python function for the amplitude bounds for every frequency.},captionpos=b, label={lst:allfreq}]
def DFT_bounds(intervalsignal): 
    N = len(intervalsignal)
    BOUNDS_I, BOUNDS_C = [],[]
    for k in range(1,N//2):
        bI,bC = AmplitudeBounds(intervalsignal, k)
        BOUNDS_I.append(bI)
        BOUNDS_C.append(bC)
    return BOUNDS_I, BOUNDS_C
\end{lstlisting}

\section{Conclusions}
We have presented three methods for computing the bounds of the Fourier amplitude spectrum when an interval signal is provided. We have shown how the problem can be tackled with a brute-force method, but only on relatively small interval signals ($<$ 20 intervals), with limited applications in practice. Despite its inefficiency the brute-force method has provided the inspiration for the second more efficient \emph{selective} method, which yields the best possible bounds on the amplitude. In this work, we have questioned the rigour and reliability of the \emph{selective} algorithm, which has led us to our third method, which uses complex interval arithmetic to propagate the bounds through the Fourier series. The third method called \emph{bounding box}, has provided a viable efficient tool to verify the rigorousness of the \emph{selective} method even for large interval signals. The \emph{bounding-box} method has also provided a robust alternative to obtain the amplitude bounds when efficiency is a priority at the expense of losing specificity in the bounds of the amplitude.

\section{Replicability}
The code, the example and the algorithms presented in this paper can be accessed in a single instance via Github at: 
\href{https://github.com/marcodeangelis/Fourier-transform}{https://github.com/marcodeangelis/Fourier-transform}.

\section{Acknowledgements}
This research is funded by the Engineering \& Physical Sciences Research Council (EPSRC) with grant no. EP/R006768/1. The research is also funded by the  Deutsche Forschungsgemeinschaft (DFG, German Research Foundation) with grant no. BE 2570/4-1 and CO 1849/1-1. EPSRC and DFG are greatly acknowledged for their funding and support. 

\bibliographystyle{rec-bst}
\bibliography{biblio_copy}

\begin{thebibliography}{12}
\expandafter\ifx\csname natexlab\endcsname\relax\def\natexlab#1{#1}\fi
\expandafter\ifx\csname bibnamefont\endcsname\relax
  \def\bibnamefont#1{#1}\fi
\expandafter\ifx\csname bibfnamefont\endcsname\relax
  \def\bibfnamefont#1{#1}\fi
\expandafter\ifx\csname citenamefont\endcsname\relax
  \def\citenamefont#1{#1}\fi
\expandafter\ifx\csname url\endcsname\relax
  \def\url#1{\texttt{#1}}\fi
\expandafter\ifx\csname urlprefix\endcsname\relax\def\urlprefix{}\fi
\providecommand{\bibinfo}[2]{#2}
\providecommand{\eprint}[1]{\href{http://arxiv.org/abs/#1}{arXiv:#1}}

\bibitem[{\citenamefont{Newland}(1993)}]{newland1993introduction}
\bibinfo{author}{\bibfnamefont{D.}~\bibnamefont{Newland}},
  \emph{\bibinfo{title}{An Introduction to Random Vibrations, Spectral and
  Wavelet Analysis}}, Dover books on engineering (\bibinfo{publisher}{Longman
  Scientific \& Technical}, \bibinfo{year}{1993}), ISBN
  \bibinfo{isbn}{9780582215849}.

\bibitem[{\citenamefont{Muscolino et~al.}(2012)\citenamefont{Muscolino,
  Santoro, and Sofi}}]{muscolino2012frequency}
\bibinfo{author}{\bibfnamefont{G.}~\bibnamefont{Muscolino}},
  \bibinfo{author}{\bibfnamefont{R.}~\bibnamefont{Santoro}}, \bibnamefont{and}
  \bibinfo{author}{\bibfnamefont{A.}~\bibnamefont{Sofi}}, in
  \emph{\bibinfo{booktitle}{5th International conference on Reliable
  Engineering Computing (REC2012). Edited by Brno, Czech Republic: Ing.
  Vladislav Pokorn{\`y}-LITERA}} (\bibinfo{year}{2012}),
  \bibinfo{pages}{407--25}.

\bibitem[{\citenamefont{Comerford et~al.}(2015)\citenamefont{Comerford,
  Kougioumtzoglou, and Beer}}]{comerford2015quantifying}
\bibinfo{author}{\bibfnamefont{L.}~\bibnamefont{Comerford}},
  \bibinfo{author}{\bibfnamefont{I.~A.} \bibnamefont{Kougioumtzoglou}},
  \bibnamefont{and} \bibinfo{author}{\bibfnamefont{M.}~\bibnamefont{Beer}},
  \bibinfo{title}{On quantifying the uncertainty of stochastic process power
  spectrum estimates subject to missing data},
  \emph{\bibinfo{journal}{International Journal of Sustainable Materials and
  Structural Systems}} \textbf{\bibinfo{volume}{2}}, \bibinfo{pages}{185}
  (\bibinfo{year}{2015}).

\bibitem[{\citenamefont{Sneddon}(1995)}]{sneddon1995fourier}
\bibinfo{author}{\bibfnamefont{I.~N.} \bibnamefont{Sneddon}},
  \emph{\bibinfo{title}{Fourier transforms}} (\bibinfo{publisher}{Courier
  Corporation}, \bibinfo{year}{1995}).

\bibitem[{\citenamefont{James}(2011)}]{james2011studentsguide}
\bibinfo{author}{\bibfnamefont{J.~F.} \bibnamefont{James}},
  \emph{\bibinfo{title}{A Student's Guide to Fourier Transforms: With
  Applications in Physics and Engineering}}, Student's Guides
  (\bibinfo{publisher}{Cambridge University Press}, \bibinfo{year}{2011}),
  \bibinfo{edition}{3rd} ed.

\bibitem[{\citenamefont{Cooley and Tukey}(1965)}]{cooley1965algorithm}
\bibinfo{author}{\bibfnamefont{J.~W.} \bibnamefont{Cooley}} \bibnamefont{and}
  \bibinfo{author}{\bibfnamefont{J.~W.} \bibnamefont{Tukey}},
  \bibinfo{title}{An algorithm for the machine calculation of complex fourier
  series}, \emph{\bibinfo{journal}{Mathematics of computation}}
  \textbf{\bibinfo{volume}{19}}, \bibinfo{pages}{297} (\bibinfo{year}{1965}).

\bibitem[{\citenamefont{Cormen et~al.}(2009)\citenamefont{Cormen, Leiserson,
  Rivest, and Stein}}]{cormen2009introduction}
\bibinfo{author}{\bibfnamefont{T.~H.} \bibnamefont{Cormen}},
  \bibinfo{author}{\bibfnamefont{C.~E.} \bibnamefont{Leiserson}},
  \bibinfo{author}{\bibfnamefont{R.~L.} \bibnamefont{Rivest}},
  \bibnamefont{and} \bibinfo{author}{\bibfnamefont{C.}~\bibnamefont{Stein}},
  \emph{\bibinfo{title}{Introduction to algorithms}} (\bibinfo{publisher}{MIT
  press}, \bibinfo{year}{2009}).

\bibitem[{\citenamefont{Stoer and Bulirsch}(2013)}]{stoer2013introduction}
\bibinfo{author}{\bibfnamefont{J.}~\bibnamefont{Stoer}} \bibnamefont{and}
  \bibinfo{author}{\bibfnamefont{R.}~\bibnamefont{Bulirsch}},
  \emph{\bibinfo{title}{Introduction to numerical analysis}},
  vol.~\bibinfo{volume}{12} (\bibinfo{publisher}{Springer Science \& Business
  Media}, \bibinfo{year}{2013}).

\bibitem[{\citenamefont{Alefeld and
  Herzberger}(2012)}]{alefeld2012introduction}
\bibinfo{author}{\bibfnamefont{G.}~\bibnamefont{Alefeld}} \bibnamefont{and}
  \bibinfo{author}{\bibfnamefont{J.}~\bibnamefont{Herzberger}},
  \emph{\bibinfo{title}{Introduction to Interval Computation}}, Computer
  Science and Applied Mathematics (\bibinfo{publisher}{Elsevier Science},
  \bibinfo{year}{2012}), ISBN \bibinfo{isbn}{9780080916361}.

\bibitem[{\citenamefont{Moore}(1966)}]{moore1966interval}
\bibinfo{author}{\bibfnamefont{R.~E.} \bibnamefont{Moore}},
  \emph{\bibinfo{title}{Interval analysis}}, vol.~\bibinfo{volume}{4}
  (\bibinfo{publisher}{Prentice-Hall Englewood Cliffs}, \bibinfo{year}{1966}).

\bibitem[{\citenamefont{Moore}(1979)}]{moore1979methods}
\bibinfo{author}{\bibfnamefont{R.~E.} \bibnamefont{Moore}},
  \emph{\bibinfo{title}{Methods and applications of interval analysis}}
  (\bibinfo{publisher}{SIAM}, \bibinfo{year}{1979}).

\bibitem[{\citenamefont{Moore et~al.}(2009)\citenamefont{Moore, Kearfott, and
  Cloud}}]{moore2009introduction}
\bibinfo{author}{\bibfnamefont{R.}~\bibnamefont{Moore}},
  \bibinfo{author}{\bibfnamefont{R.}~\bibnamefont{Kearfott}}, \bibnamefont{and}
  \bibinfo{author}{\bibfnamefont{M.}~\bibnamefont{Cloud}},
  \emph{\bibinfo{title}{Introduction to Interval Analysis}}, Other titles in
  applied mathematics (\bibinfo{publisher}{Cambridge University Press},
  \bibinfo{year}{2009}), ISBN \bibinfo{isbn}{9780898716696}.

\end{thebibliography}

\end{document}